\documentstyle[preprint,prb,aps]{revtex}

\tighten

\begin{document}

\draft

\preprint{MA/UC3M/16/94}

\title{Self-consistent analysis of electric field effects on 
Si $\delta$-doped GaAs}

\author{Jos\'e A.\ Cuesta and Angel S\'anchez}

\address{Escuela Polit\'ecnica Superior,
Universidad Carlos III de Madrid,
C./ Butarque 15,  E-28911 Legan\'es, Madrid, Spain}

\author{Francisco Dom\'{\i}nguez-Adame}

\address{Departamento de F\'{\i}sica de Materiales,
Facultad de F\'{\i}sicas, Universidad Complutense,
E-28040 Madrid, Spain}

\maketitle

\begin{abstract}

We theoretically study the subband structure of single Si $\delta$-doped
GaAs inserted in a quantum well and subject to an electric field applied
along the growth direction. We use an efficient self-consistent procedure
to solve simultaneously the Schr\"odinger and Poisson equations for 
different values of electric field and temperature. We thus find the 
confining potential, the subband energies and their corresponding envelope 
functions, the subband occupations, and the oscillator strength of 
intersubband transitions. Opposite to what is usually the case when 
dealing with the quantum-confined Stark effect in ordinary quantum wells, 
we observe an abrupt drop of the energy levels whenever the external field 
reaches a certain value. This critical value of the field is seen to
depend only slightly on temperature. The rapid change in the energy levels 
is accompanied by the appearance of a secondary well in the confining 
potential and a strong decrease of the oscillator strength between the 
two lowest subbands. These results open the possibility to design devices
for use as optical filters controlled by an applied electric field. 

\end{abstract}


\begin{multicols}{2}

\narrowtext

\section{Introduction}

Recent advances in epitaxial growth techniques, such as molecular beam
epitaxy (MBE), make possible 
to fabricate high quality Si $\delta$-doped layers in GaAs. 
In these systems, a slab of Si atoms localized within few
monolayers supplies electrons and gives rise to quantum confinement of
carriers. By this means, a two-dimensional electron gas can be realized by
planar doping of GaAs at high donor
concentration.\cite{Zrenner} Devices based 
on $\delta$-doped heterostructures are currently under extensive 
investigation for high-speed electronic and optoelectronic applications
(see, e.g., Ref.\ \onlinecite{Wall} for some examples of the practical 
advantages of $\delta$-doping). Hence the interest of a good 
understanding of Si $\delta$-doped GaAs as a representative example of 
those devices.

Theoretical studies of the above systems usually neglect possible effects of 
disorder due to the random distribution of impurities in order to simplify 
the analysis. 
Indeed, currently available techniques allow for an optimal control of the 
growing heterostructure, thus justifying the
assumption that the ionized impurity atoms are homogeneously 
distributed inside the $\delta$-doped layers. This approximation has been 
recently shown to be correct in the high-density limit.\cite{Kortus}  
A number of researches have considered this limit within different 
approaches, like the Thomas-Fermi,\cite{Ioratti} local density 
approximation (LDA), \cite{Degani} and Hartree methods.\cite{Zrenner2}  
These previous works show that in the absence of external fields the 
Thomas-Fermi semiclassical approach is equivalent to a self-consistent 
formulation in a wide range of doping concentrations.\cite{Ioratti}  
The effects of applied 
electric field have recently been considered in the case of 
single and periodically Si $\delta$-doped GaAs \cite{SST,PRB} by using a
generalized Thomas-Fermi formalism. In this framework one first computes 
the one-electron potential in the absence of electric field and then
tilts it to account for the electric potential. It has thus been found 
that the well-known Stark ladders, already observed in quantum-well based 
superlattices,\cite{Agullo} should also be clearly revealed in 
periodically Si $\delta$-doped GaAs. Nevertheless, this {\em ad hoc} 
procedure might be incapable to describe all the phenomenology arising in
$\delta$-doped heterostructures: A fully self-consistent approach may 
reveal interesting issues about the behavior of actual structures 
under bias voltage not accounted for within simpler approaches. 

As an example of new phenomenology found by self-consistent procedures,
we mention that, in superlattice-like systems, the Hartree potential from 
electrons of different wells partly screens the effect of the electric 
field leading to the formation of electric field domains. The physical 
existence of electric field domains is firmly established in 
GaAs-Ga$_{1-x}$Al$_{x}$As \cite{Grahn} superlattices 
(see Ref.\ \onlinecite{Bonilla} and references therein). To our knowledge, 
however, there have been no reports
in the literature about field domains in $\delta$-doped GaAs 
superlattices. Note that in this case there is a number of free carriers 
in the structure and, as a consequence, it is possible for the electric 
field to break up into two or more regions with different field 
strengths, i.e., electric field domains. Similarly, it would not be 
strange to observe other unexpected phenomena when a 
more complete theoretical analysis of $\delta$-doped systems 
under applied fields is carried 
out. It is therefore clear that a careful self-consistent analysis of 
this kind of devices is necessary to make sure what their properties
as well as their possible applications are. 

The work we report on in this paper is a first step in the aforementioned
direction. We have concerned ourselves with the self-consistent study of 
a single Si $\delta$-doped layer in GaAs under an applied electric field.
The aim of addressing this question is twofold. On the one hand, we 
intend to implement and improve an efficient self-consistent procedure to
analyze this simple structure, thus developing the necessary skills to 
face the full superlattice problem. It is important to realize that a 
self-consistent study of a complex heterostructure requires a well 
developed technique to prevent expensive calculations. 
On the other hand, we believe that only after reaching a thorough 
knowledge of the phenomena appearing in a single layer it will be 
possible to proceed further, in order to asses the electronic structure 
of periodically Si $\delta$-doped GaAs superlattices. 
In addition, if our work showed new features
of the single layer problem, we could foresee the subsequent, unexpected 
phenomena which would be likely to arise in the superlattice, as well as 
estimate the range of parameters for which they might take place. 
Therefore, a 
complete understanding of charge distribution and subband energy 
dependence on the applied electric field in a single Si $\delta$-doped 
GaAs layer is necessary. With 
this double goal in mind, we undertook the study of the quantum confined 
Stark effect in these structures by considering a $\delta$-doping
layer in a quantum well.

The paper is organized as follows. In Sec.\ II we briefly discuss our 
model, in which we use a scalar Hamiltonian within the effective-mass
approximation to describe the electron dynamics.  The one-electron
potential due to the combined action of the ionized donors in the single
$\delta$-doped layer and the applied electric field is found by 
simultaneously solving the Schr\"odinger and Poisson equations.  Section 
III is devoted to a summary of the numerical method we apply to obtain (i) 
the one-electron potential, (ii) the subband structure dependence upon the
applied electric field, (iii) the subband occupation as a function of the
field, (iv) the spatial charge distribution, and (v) the 
oscillator strength for
intersubband transitions when the whole structure is confined between
two infinitely high barriers in an applied electric field (quantum
confined Stark effect).  Results and discussions are collected in
Sec.\ IV and, finally, Sec.\ V ends the paper with a brief recollection
of results and some comments on possible physical consequences and 
technological applications of our results in new devices.

\section{Model}

The system we study in this work is a semiconductor structure made of a
single Si $\delta$-doped GaAs layer.  We consider a slab of GaAs of
thickness $L$ confined between two infinitely high barriers with a Si
$\delta$-doping layer embedded in its center.  We assume that the doping
layer consists of a continuous slab of thickness $d$ with $N_D^{+}$
ionized donors per unit area.  In what follows we neglect the
unintentional $p$-type background doping appearing in most MBE-grown
samples. This is not a serious limitation since actual
techniques can keep this doping level very low (less than $10^{14}\,
$cm$^{-3}$ acceptors). For such a low residual doping, the well shape
below the Fermi level $E_F$ is almost insensitive to the background
acceptors.\cite{Johnson}

We assume the validity of the effective-mass approximation and take an
isotropic and parabolic conduction-band in the growth direction.  This
approximation usually works fine in GaAs, except a very high electric
fields, when $\Gamma-X$ mixing induced by the field occurs.
\cite{Hagon}  Kane's parameter ($E_p\sim 23\,$meV) and the
conduction-band modulation are much smaller than the bandgap in
GaAs, and the coupling between host bands is small, so that a scalar
Hamiltonian suffices to properly describe the electronic conduction-band
states in $\delta$-doped GaAs.\cite{SST2}  In the envelope function
approach, the electronic wave function corresponding to the $j$-subband
may be factorized as follows
\begin{equation}
\psi_j({\bf r})={1\over \sqrt{S}}\,\exp(i{\bf k}_\perp\cdot {\bf
r}_\perp)\psi_j(z),
\label{eq1}
\end{equation}
where ${\bf k}_\perp$ and ${\bf r}_\perp$ are the in-plane wavevector
and spatial coordinates, respectively.  Here $S$ is the area of the
layer.  The subband energy follows the parabolic dispersion
law $E_j+\hbar^2k_\perp^2/2m^*$, $m^*$ being the electron
effective-mass at the bottom of the conduction band ($\Gamma$ valley).
The quantized energy levels $E_j$ and their corresponding envelope
functions $\psi_j(z)$ satisfy the following Schr\"odinger-like equation
\begin{equation}
\left[-\,{\hbar^2\over 2 m^*}\,{d^2\phantom{z}\over dz^2}+V(z)\right]\,
\psi_j(z)=E_j\psi_j(z),
\label{eq2}
\end{equation}
The presence of infinite barriers implies that the envelope functions
vanish at $z=\pm L/2$.

The one-electron potential splits into four different contributions
\begin{equation}
V(z)=V_b(z)+eFz+V_{xc}(z)+V_H(z).
\label{eq3}
\end{equation}
Here $V_b(z)$ is the built-in potential due to the infinite barriers.
Therefore $V_b(z)=0$ for $|z|<L/2$ and becomes infinite otherwise. $F$
is the strength of the applied electric field. $V_{xc}(z)$ is the local
exchange-correlation energy calculated in the LDA approximation using
the Hedin-Lundqvist parametrization \cite{Hedin}
\begin{equation}
V_{xc}(z)=-\,{Ry^*\over 10.5\pi\alpha r}\,[1+0.7734\,r\ln(1+1/r)],
\label{eq4}
\end{equation}
where we have defined
$$\alpha=(4/9\pi)^{1/3}\ \ \ \ \ \mbox{and}\ \ \ \ \
r={1\over 21}\left({4\over 3}\pi a^{*3}n(z)\right)^{-1/3},
$$
$n(z)$ is the electron density, and $Ry^*=
e^2/(8\pi\epsilon_0\kappa a^*)$ and $a^*=4\pi\epsilon_0\kappa\hbar^2/
m^*e^2$ are respectively
the effective Rydberg and the effective Bohr radius.  The
local dielectric constant $\kappa$ is assumed not to depend on
the spatial coordinate in the
whole structure.  The Hartree potential $V_H(z)$ is obtained by solving
the one-dimensional Poisson equation
\begin{equation}
{d^2V_H(z)\over dz^2}={e\over\epsilon_0\kappa}\,[N^{3D}_D(z)-n(z)],
\label{eq5}
\end{equation}
along with the boundary conditions $V_H(z=-L/2)=V_H(z=+L/2)=0$, where
$N^{3D}_D(z)=N_D^{+}/d$ for $|z|<d/2$ and vanishes in other regions.
The electron density can be written as
\begin{equation}
n(z)=\sum_j\>n_j|\psi_j(z)|^2,
\label{eq6}
\end{equation}
where the sum runs over the different subbands. The subband occupation
$n_j$ is given by
\begin{equation}
n_j={k_BTm^*\over \pi\hbar^2}\,\ln \left[ 1+\exp\left( {E_F-E_j \over
k_BT} \right) \right].
\label{eq7}
\end{equation}
This set of equations must be solved until self-consistency is reached.
Then, envelope functions and electron energies can be found as a
function of the applied field.  Once these magnitudes are computed, it
is easy to obtain the charge distribution and subband occupations. In
addition, regarding field-dependent intersubband transitions from a
state $k$ into a state $j$, it becomes most important to determine the
corresponding oscillator strength
\begin{equation}
f_{jk}={2m_0(E_k-E_j)\over\hbar^2}\,|\langle\psi_k|z|\psi_j\rangle|^2,
\label{eq8}
\end{equation}
where $m_0$ is the free-electron mass. Device applications of
intersubband transitions require a large Stark shift with a high
oscillator strength. For example, it is known that square quantum
wells exhibit a weak intersubband transition \cite{Harwit} but it can
be improved by inserting one narrow well inside a wider one.
\cite{Yuh,Chen} Therefore it is interesting, from a technological
viewpoint, to compare the oscillator strength in $\delta$-doped
structures with typical values obtained in quantum wells.

\section{Numerical analysis}

\subsection{Adimensionalization}

For convenience, we begin by introducing
dimensionless variables in our problem. We can
define reduced well length, $l$, doping layer thickness, $\Delta$, and
position coordinate, $x$, by rescaling $L$, $d$, and $z$ with $a^*$,
respectively. Further, we introduce reduced energies, $\epsilon$, 
and potentials, $v$, by rescaling $E$ and $V$ 
with $Ry^*$. On the other hand, reduced electric fields,
$f$, charge densities, $\nu$, and temperatures, $\tau$, are defined by
rescaling $F$, $n$, and $T$ respectively with their corresponding scaling
factors
$F_0\equiv Ry^*/(ea^*)\approx 5.91$ kV cm$^{-1}$, $n_0\equiv
(a^*)^{-3}\approx 10^{18}$ cm$^{-3}$, and $T_0\equiv Ry^*/k_B\approx
67.7$ K. With these new variables Eq.\ (\ref{eq3}) becomes
\begin{equation}
v(x)=v_b(x)+fx+v_{xc}(x)+v_H(x).
\label{eq9}
\end{equation}
In this equation, 
$v_b(x)=0$ for $|x|<l/2$ and is infinite otherwise. The reduced
exchange potential, $v_{xc}$, is given by
\begin{mathletters}
\begin{eqnarray}
v_{xc}(x) & = & -\,\frac{1}{7}\left(\frac{2}{3\pi^2}\right)
\left\{u(x)+0.7734\ln\left[1+u(x)\right]\right\},
\label{eq10a}  \\
u(x) & \equiv & 7(36\pi)^{1/3}\nu(x)^{1/3},
\label{eq10b}
\end{eqnarray}
\end{mathletters}
and $v_H(x)$ is determined by solving
\begin{equation}
v_H''(x)=8\pi\left[\nu_D^{3D}(x)-\nu(x)\right].
\label{eq11}
\end{equation}
Here $\nu_D^{3D}(x)=(a^*)^2N_D^+/\Delta$ for $|x|<\Delta/2$ and 0
otherwise, and $\nu(x)$ is computed through
\begin{mathletters}
\begin{eqnarray}
\nu(x) & = & \sum_{j}\nu_j|\chi_j(x)|^2,
\label{eq12a}  \\
\nu_j & = & \frac{\tau}{2\pi}\ln\left[1+\exp\left(\frac{\epsilon_F-
\epsilon_j}{\tau}\right)\right],
\label{eq12b}
\end{eqnarray}
\end{mathletters}
where $\chi_j(x)$ are the eigenfunctions of the Schr\"odinger problem
\begin{equation}
-\chi_j''(x)+v(x)\chi(x)=\epsilon_j\chi_j(x)
\label{eq13}
\end{equation}
with boundary conditions $\chi_j(\pm l/2)=0$.

\subsection{Discretization}

The numerical procedure requires discretizing the $x$ variable as
$x_i\equiv ih-l/2$, with $h\equiv l/(N+1)$, for $i=0,1,\dots,N+1$.
Equation (\ref{eq13}) can then be approximated by the following
difference equation
\begin{eqnarray}
-\frac{1}{h^2}\left[\chi_j(x_{i+1})-2\chi_j(x_i)+\chi_j(x_{i-1})\right]
 & + & \nonumber \\
 + v(x_i)\chi_j(x_i)&=&\epsilon_j\chi_j(x_i),
\label{eq14}
\end{eqnarray}
with boundary conditions $\chi_j(x_0)=\chi_j(x_{N+1})=0$. The
problem above is nothing but the diagonalization of the symmetric,
tridiagonal $N\times N$ matrix $H$, defined as
\begin{equation}
H_{ij}\equiv\left\{\begin{array}{ll}
v(x_i)+2h^{-2} & \mbox{ if $i=j$ } \\
-h^{-2} & \mbox{ if $|i-j|=1$ } \\
0 & \mbox{ otherwise }
\end{array}, \right.
\label{eq15}
\end{equation}
$\epsilon_j$ being the $j$-th eigenvalue and $\chi_j(x_i)$
($i=1,2,\dots,N$) the $i$-th component of the $j$-th eigenvector, with
$j$ running from 1 to $N$. The Schr\"{o}dinger equation is thus
transformed into the much simpler problem of diagonalizing $H$, and
we can take advantage of its simpler shape.

Once $\epsilon_j$ and $\chi_j$ are available for $j=1,2,\dots,N$, the
Fermi level, $\epsilon_F$, is obtained as the solution of
\begin{equation}
(a^*)^2N_D^+=\sum_{j=1}^N\nu_j,
\label{eq16}
\end{equation}
with $\nu_j$ given by (\ref{eq12b}) for every $\epsilon_F$. Now the
electron density $\nu(x_i)$ is completely determined via Eq.\
(\ref{eq12a}) and, accordingly, the full right hand side of Eq.\
(\ref{eq11}). Again this latter equation can be approximated by a
difference equation, namely
\begin{eqnarray}
h^{-2}\left[v_H(x_{i+1})-2v_H(x_i)+v_H(x_{i-1})\right]
 = & & 
\nonumber \\
  \phantom{sitio} = 8\pi  \left[\nu_D^{3D}(x_i)-\nu(x_i)\right], & &
\label{eq17}
\end{eqnarray}
with boundary conditions $v_H(x_0)=v_H(x_{N+1})=0$. 
This problem can be straightforwardly mapped onto 
that of finding the solution to the system $Dv_H=\rho$, where $D$
is the symmetric, tridiagonal $N\times N$ matrix defined as
\begin{equation}
D_{ij}\equiv\left\{\begin{array}{ll}
-2 & \mbox{ if $i=j$ } \\
1 & \mbox{ if $|i-j|=1$ } \\
0 & \mbox{ otherwise }
\end{array}, \right.
\label{eq18}
\end{equation}
and $\rho$ is the vector whose components are $\rho_i\equiv 8\pi h^2
\left[\nu_D^{3D}(x_i)-\nu(x_i)\right]$, $i=1,2,\dots,N$. The solution,
$v_H$, of this linear system can be readily obtained through the
standard Thomas's algorithm, which amounts to computing the solution
iteratively, as follows:
\begin{equation}
\begin{array}{rcl}
v_H(x_{N+1}) & = & 0,  \\
v_H(x_j) & = & \alpha_jv_H(x_{j+1})+\gamma_j,
\end{array}
\label{eq19}
\end{equation}
with the index $j$ running backwards ($j  =  N,N-1,\dots,1$), and where
\begin{equation}
\begin{array}{rclrcl}
\alpha_0 & = & 0, & \gamma_0 & = & 0, \\
 & & & & & \\
\alpha_j & = & {\displaystyle \frac{-c_j}{d_j+a_j\alpha_{j-1}} }, &
\gamma_j & = & {\displaystyle \frac{b_j-a_j\gamma_{j-1}}{d_j+
a_j\alpha_{j-1}} },\\
 & & & & & \\
j & = & 1,2,\dots,N,  
\end{array}
\label{eq20}
\end{equation}
$d_j$ ($j=1,2,\dots,N$), $a_j$ ($j=2,3,\dots,N$), and $c_j$
($j=1,2,\dots,N-1$) being the diagonal, subdiagonal, and superdiagonal
elements of the matrix $D$
respectively (in our case, $d_j=-2$ and $a_j=c_j=1$). Note that the 
undefined elements $a_1$ and $c_N$ appearing in Eq.\ (\ref{eq20}) are
in fact irrelevant. 

\subsection{Algorithm}

The self-consistent algorithm consists of the following steps:
\begin{enumerate}
\item Set $v_H(x_i)=\nu(x_i)=0$ for all $i=1,2,\dots,N$.
\item Set $v_H^{\rm old}=v_H$.
\item Compute $v(x_i)$ as given by Eq.\ (\ref{eq9}).
\item Diagonalize $H$ [Eq.\ (\ref{eq15})] to obtain the eigenvalues,
$\epsilon_j$, and eigenvectors, $\chi_j$ ($j=1,2,\dots,N$).
\item Compute $\epsilon_F$ by solving Eq.\ (\ref{eq16}).
\item Compute $\nu(x_i)$ as given by Eqs.\ (\ref{eq12a}--\ref{eq12b}).
\item Determine $v_H^{\rm new}(x_i)$ through the recurrence
(\ref{eq19}) (notice that the coefficients $\alpha_j$ and $\gamma_j$
only have to be computed once at the beginning of the program).
\item Check for convergence (by comparing, for instance, $v_H^{\rm new}$
with $v_H^{\rm old}$); if tolerance has been attained then exit the
self-consistence process with $v_H=v_H^{\rm new}$.
\item Otherwise set $v_H=\lambda v_H^{\rm new}+(1-\lambda)v_H^{\rm old}$
and repeat from step 2 on.
\end{enumerate}
The parameter $\lambda$ has been introduced to control the iteration: if
$\lambda=0$ we will have $v_H=0$ forever, while if $\lambda=1$, $v_H$
will undergo its maximum variation at every step. This latter case
(commonly used throughout literature) has proven not to converge for
$N_D^+\gtrsim 10^{-12}$ cm$^{-2}$. Instead, $\lambda=1/2$ makes
the process convergent for any set of parameters.

The numerical parameters we have been using are as follows. Our mesh
consisted of 501 points, enough as to resolve the physical dimensions
of the delta layer potential we will describe below. We have checked
our numerical procedure by using smaller and larger numbers of grid 
points, finding a negligible dependence on this parameter, as desired. We 
chose to stop the iterative procedure when the relative variation of
the Fermi energy was $\leq
10^{-6}$. We also used another criterium, namely
computing the integral of the 
absolute value of the difference between the new and old 
potentials and stopping when it was less than $10^{-3}$, obtaining 
again good agreement between both conditions. Typical runs attained
convergence after about 20 Hartree iterations, each one of them taking
around 30 s of CPU time on a HP 9000/715/75 with the above integration 
parameters. This shows that our procedure is rather efficient, and we
are confident that it will allow us to deal with the superlattice 
problem without prohibitive CPU time requirements. 

\section{Results and discussions}

The different magnitudes we are interested in have been obtained taking
$m^*=0.067m_0$ and $\kappa=12.7$ in GaAs.  Since the dependence of the
subband structure upon donor concentration ($N_D^*$), its distribution
width ($d$) and the thickness ($L$) are well understood in the absence
of external fields,\cite{Degani2} even if the whole structure is
embedded in a quantum well,\cite{Xu} we have fixed their values and
concerned ourselves with the dependence on temperature and electric field.
In our numerical simulations we 
have set $N_D^+=5\times 10^{12}\,$cm$^{-2}$,
$d=20\,$\AA\ and $L=500\,$\AA. This choice, with the above numerical 
parameters, gives our calculation a resolution of 1\AA.
The maximum electric field we have
considered is $F=100\,$kV cm$^{-1}$, 
well below the value for which $\Gamma-X$
mixing may be observed, thus remaining within the validity range
of our scalar 
Hamiltonian.  We have
studied three typical temperatures, namely those of liquid 
helium, $4.2\,$K, liquid nitrogen, $77\,$K, and room temperature,
$300\,$K.

In the absence of applied electric field, the confining potential
presents the characteristic V-shape profile, as shown in
Fig.\ \ref{fig1}, where the origin of energies of all curves is the
Fermi level.  On increasing the electric field the potential is tilted
so that it becomes slightly asymmetric (see Fig.\
\ref{fig1} for $50\,$kV\,cm$^{-1}$).  This behavior is also
predicted by means of the Thomas-Fermi approach, as mentioned above.
\cite{SST}  This trend holds until a critical value of the electric
field $F_c$ (about $60\,$kV\,cm$^{-1}$, see below) is reached.  For
higher strengths of the electric field, the confining potential changes
dramatically its shape and a second minimum appears at the left barrier.
Therefore, a local potential
barrier arises between the center and the left
barrier of the structure, as shown in Fig.\ \ref{fig1} for
$F=100\,$kV\,cm$^{-1}$.  It is worth mentioning that the transition
between these two different regimes is very sharp.  In other words,
$V(-L/2)$ drops $\sim 300\,$meV in a very narrow 
interval, $\leq 1 kV cm^{-1}$, of electric fields.
This phenomenon can be thought of as arising from competition between 
opposite effects: On one side, the electric field pushes the charge 
distribution to the left, and on the other side, electronic repulsion
prevents charge accumulation on that part of the system. 

The different shapes of the confining potential for $F$ smaller or
larger than $F_c$ must strongly influence the subband structure and the
corresponding subband occupation, as it actually occurs.  Results at
different temperatures are collected in Fig.\ \ref{fig2}.  In this
figure we have 
only plotted those subbands whose electron densities $n_j$ are
at least $1\%$ of the ground state occupation. As expected, the
occupation of subbands above the ground state
increases with temperature.  The
subband energies depend only slightly on the applied electric field
up to a critical field $F_c\approx 60\,$kV\,cm$^{-1}$, at which a
sudden drop of the levels is observed.  This value of $F_c$ is almost
the same for $T=4.2\,$K and $T=77\,$K, whereas at room temperature is
somewhat smaller.  It is clear that this drop of the levels is related
with the rapid change in the value of $V(-L/2)$ and the subsequent
appearance of the local maximum around $z=-L/4$.  This behavior is
very different to the quantum confined Stark effect in quantum wells,
where subbands change smoothly with $F$.\cite{Matsuura}  The subband
occupations at three different temperatures and three different values of
the electric field are shown in Table~\ref{table1}.  The occupation of the
lowest subband is almost independent of $T$ and $F$.  On increasing
temperature, more and more subbands are populated, as mentioned
above.  The most significant
feature when increasing the electric field is that the occupation of the
excited subbands increases rapidly for fields above $F_c$.  This fact is
remarkable at low temperature, for which the first excited state has an
occupation of $0.007$ (in units of $10^{12}\,$cm$^{-2}$) for $F<F_c$
while it increases two orders of magnitude for $F>F_c$.  Hence band
filling also changes dramatically when crossing $F_c$.

According to our previous results, the spatial charge distribution
should also be strongly influenced by the applied electric field.  To
elucidate these effects, we have calculated the squared envelope
functions at low temperature (Fig.\ \ref{fig3}).  At zero
field the envelope functions present a well defined parity since $V(z)$
is an even function; thus the electron density $n(z)$ is symmetric
around the $\delta$-doping layer.  On increasing the electric field, the
expected value $\langle\psi_j |z|\psi_j\rangle$ shifts to the left, in
the same fashion as in ordinary quantum wells.  However, electric fields
larger than $F_c$ cause the first excited state to be located very close
to the left barrier due to the presence of the potential minimum at
$z=-L/2$.  Therefore, spatial charge distribution undergo a large shift
to the left for $F>F_c$, which must clearly have profound effects on
intersubband transitions between the first excited subband and the ground
subband.  This is shown in Fig.\ \ref{fig4}, where $f_{10}$ is plotted
as a function of the electric field at three different temperatures.
Below $F_c$ the oscillator strength is almost constant and presents a
high value, close to $15$, which is even larger than that obtained for a
narrow GaAs-Ga$_{1-x}$Al$_{x}$As quantum well inside a wider one.
\cite{Chen}  Thus, for instance, $f_{10}(50\,$kV\,cm$^{-1}) /
f_{10}(0\,$kV\,cm$^{-1})\sim 0.93$.  But contrary to quantum wells,
$f_{10}$ decays very fast for electric fields larger than $F_c$,
so that, for instance, 
$f_{10}(100\,$kV\,cm$^{-1}) / f_{10}(0\,$kV\,cm$^{-1})\sim 0.1$.
This change is practically the same 
for the two lowest temperatures, the magnitude of the jump becoming 
smaller at room temperature; nevertheless,
the drop of $f_{10}$ is equally
abrupt for all temperatures. In closing this section, we note that this
drop of the oscillator strength arises from the fact that the secondary
well, newly formed, begins to confine charge. As a result, the overlapping
between the ground and the first excited states is small, and transitions
between both subbands become tunneling-like, leading to a sharp decrease of
$f_{10}$. 

\section{Conclusions}

In the present work we have studied single Si $\delta$-doped GaAs under
an electric field applied parallel to the growth direction.  Electronic
structure and intersubband transitions have been calculated by solving
the Schr\"odinger and Poisson equations self-consistently. To this end,
we have developed a very efficient numerical code which paves the 
way to self-consistent studies of $\delta$-doped superlattices. As 
regards the specific system we deal with here, one of the
most significant results is the existence of a critical value of the
electric field $F_c$ separating two very different behaviors of the
quantum confined Stark effect, which is very different from 
what happens in quantum wells.  This
critical field is related to the fact that the confining potential
develops a second minimum at the left barrier.  As a consequence, we
have observed that the subband energies present a step-like shape as a
function of the electric field, with a sharp decrease at $F_c$.  This
is accompanied by a  marked increase of the occupation of the first
excited subband and a strong shift of the spatial charge distribution.
Finally, the oscillator strength for intersubband transitions is very
large and almost constant below $F_c$ but vanishes quickly above $F_c$.

Interestingly, this sharp change in the confining potential shape poses
a number of questions about the behavior of superlattices. From what 
we now know, it is not clear whether the appearance of an intermediate
barrier will help form electric field domains or, on the contrary, will 
prevent them due to loss of coherence. It is conceivable that electric
field domains will appear below $F_c$, where the phenomenology is 
similar to that of quantum-well-based superlattices. The situation is more
intriguing above $F_c$, because the splitting of the quantum well into two
wells might or might not survive in the superlattice. The consequences of
the preservation of this effect are difficult to predict, as in that case
the superlattice might exhibit a double periodicity or become aperiodic. 
If double periodicity arises, it could induce unusual electron transport
properties, like those predicted for the so-called dimer superlattices.
\cite{remoco}
In addition, it is also evident that the well splitting phenomenon
will have consequences as to possible resonant tunneling effects, not 
only in superlattices, but also in double $\delta$-doped layer structures.
These and related questions are clearly 
worth studying, and work along these lines is now in progress.\cite{eh}

Finally, we
want to end the paper with a few comments on applications of what
we have found. As to applications, 
the fact that the oscillator strength undergoes a sudden drop allows
us to suggest that structures made out of 
Si $\delta$-doped GaAs embedded in a quantum well 
may be of use as optical
filters. Note that dipolar transitions $1\to 0$ show a negligible
probability above $F_c$; as these two subbands contain most of the 
carriers, this will be the relevant transition governing the optical 
response of the device.  It is important to recall that the value of
$F_c$ is almost independent on temperature, so that such a device could
be used in a wide range of temperatures.  In addition, it is clear that
the value of $F_c$ could be adjusted by changing $N_D^+$, $L$ and $d$.
Therefore, this kind of structure may open the possibility of
fabricating new optoelectronic devices. 

\acknowledgments

J.\ A.\ C.\ thanks partial 
financial support from DGICyT (Spain) through project
PB91-0378.
Work at Legan\'es is supported by the DGICyT (Spain) through project
PB92-0248, and by the European Union Human Capital and Mobility
Programme through contract ERBCHRXCT930413.  Work at Madrid is supported
by UCM through project PR161/93-4811.

\newpage 

\begin{table}
\caption{The subband occupation for three different temperatures and
electric fields is shown in units of $10^{12}\,$cm$^{-2}$.}
\label{table1}
\end{table}

\begin{tabular}{|c|c|c|c|c|c|c|} \hline 
\rule[-.25cm]{0cm}{.7cm}
T (K) & F (kV cm$^{-1}$) & Ground & 1st & 2nd & 3rd & 4th \\ \hline
\hline
\rule[-.25cm]{0cm}{.7cm}
 & 0 & 2.50 & 0.007 & 0 & 0 & 0 \\ \cline{2-7}
\rule[-.25cm]{0cm}{.7cm}
4.2 & 50 & 2.42 & 0.007 & 0 & 0 & 0 \\ \cline{2-7}
\rule[-.25cm]{0cm}{.7cm}
 & 100 & 2.80 & 1.36 & 0 & 0 & 0 \\ \hline \hline
\rule[-.25cm]{0cm}{.7cm}
 & 0 & 2.48 & 0.13 & 0 & 0 & 0
 \\ \cline{2-7} 
\rule[-.25cm]{0cm}{.7cm}
77 & 50 & 2.40 & 0.13 & 0 & 0 & 
0  \\ \cline{2-7}
\rule[-.25cm]{0cm}{.7cm}
 & 100 & 2.78 & 1.34 & 0.79 & 0.05 & 0 \\ \hline \hline
\rule[-.25cm]{0cm}{.7cm}
  & 0 & 2.42 & 0.50 & 0.10 & 0.02 & 0.005 \\ \cline{2-7}
\rule[-.25cm]{0cm}{.7cm}
  300 & 50 & 2.35 & 0.50 & 0.11 & 0.03 & 0.006 \\ \cline{2-7}
\rule[-.25cm]{0cm}{.7cm}
  & 100 & 2.58 & 1.20 & 0.80 & 0.29 & 0.08 \\ \hline
\end{tabular}
 
\begin{figure}
\caption{The self-consistent confining potential for a ionized donor
concentration $N_D^+=5\times 10^{12}\,$cm$^{-2}$ distributed over
$d=20\,$\AA\ for three different electric fields, at $T=4.2\,$K. The
zero of energy is set at the Fermi level.}
\label{fig1}
\end{figure}

\begin{figure}
\caption{Subband energies as a function of the applied electric field
for (a) $T=4.2\,$K, (b) $T=77\,$K, and (c) $T=300\,$K. Only those
subbands with electron densities larger than $1\%$ of the electron
density in the ground state are plotted. Energies are referred to the
Fermi level. Note that the vertical scale is different in plot (c)
for the sake of clarity.}
\label{fig2}
\end{figure}

\begin{figure}
\caption{Squared envelope fuctions for the three lowest subbands
at different fields (a) $F=0\,$kV\,cm$^{-1}$, (b) $F=50\,$kV\,cm$^{-1}$,
and (c) $F=100\,$kV\,cm$^{-1}$, at $T=4.2$\,K.}
\label{fig3}
\end{figure}

\begin{figure}
\caption{Oscillator strength for the intersubband transition $1\to 0$ as
a function of the applied electric field at different temperatures.}
\label{fig4}
\end{figure}

\end{multicols}

\end{document}